\documentclass[secnumarabic, graphics,floatfix,tightenlines,nobibnotes, aps, prb, twocolumn]{revtex4-1}
\usepackage{amsmath,amssymb}
\usepackage{graphicx}
\usepackage{dcolumn}
\usepackage{bm}
\usepackage{braket}
\usepackage{verbatim}
\usepackage{float}
\usepackage{bbold}
\usepackage{color}
\usepackage{xcolor}
\usepackage{relsize}

\usepackage{slashed}
\usepackage{amsthm}
\usepackage[colorlinks=true ,urlcolor=blue,urlbordercolor={0 1 1}]{hyperref}

\begin{document}

\title{Vortex spin liquid with fractional quantum spin Hall effect in moir\'e Chern bands}
\author{%
Ya-Hui Zhang
}
\affiliation{William H. Miller III Department of Physics and Astronomy, Johns Hopkins University, Baltimore, Maryland, 21218, USA}

\date{\today}

\begin{abstract}
Integer and fractional quantum anomalous Hall (QAH) effects have been widely seen in moir\'e systems. Recently there is even observation of a time reversal invariant fractional quantum spin hall (FQSH) state at filling $n=3$ in twisted MoTe$_2$ bilayer.     We consider a pair of half-filled $C=\pm 1$ Chern band in the two valleys, similar to the well-studied quantum Hall bilayer, but now with opposite chiralities. Due to the strong inter-valley repulsion, we expect a charge gap opening with low energy physics dominated by the neutral inter-valley excitons. However, the presence of an effective `flux' frustrates exciton condensation by proliferating vortices. Here we construct a  vortex liquid of excitons dubbed as vortex spin liquid (VSL), from exciton pairing of the composite fermions in the decoupled composite Fermi liquids (CFL) phase. This insulator is a  quantum spin liquid with gapless spin excitations carried by the flux of an emergent U(1) gauge field. Additionally, there exist neutral and spinless Fermi surfaces formed by fermionic vortices of a nearby inter-valley-coherent (IVC) order.  Unlike a conventional Mott insulator,  the VSL phase also exhibits FQSH effect with gapless helical charge modes along the edge. Our work suggests a new platform to search for  quantum spin liquid enriched by fractional quantum spin Hall effect. We also point out the possibility of quantum oscillations and thermal Hall effect under Zeeman field in this exotic insulator.
\end{abstract}

\pacs{Valid PACS appear here}
\maketitle

\textbf{Introduction} Experimental observations of fractional chern insulator\cite{sun2011nearly,sheng2011fractional,neupert2011fractional,wang2011fractional,tang2011high,regnault2011fractional,bergholtz2013topological,parameswaran2013fractional,PhysRevB.84.165107} at zero magnetic field in twisted MoTe2 \cite{cai2023signatures,zeng2023integer,park2023observation,PhysRevX.13.031037} and in  pentalayer  rhombohedrally stacked graphene aligned with hBN\cite{2023arXiv230917436L} are one of the most important breakthroughs in condensed matter physics in the recent years. Theoretically quantum anomalous Hall (QAH) and fractional quantum anomalous Hall(FQAH) phases were expected in the moir\'e systems with nearly-flat Chern bands following spontaneous valley polarization\cite{zhang2019nearly,repellin2020ferromagnetism}. For the twisted MoTe$_2$ system with the twist angle $\theta\approx 3.7^\circ$, it is now widely believed that the first moir\'e band mimics the lowest Landau level quite well and thus supports the familiar quantum Hall physics \cite{wu2019topological,yu2020giant,devakul2021magic,li2021spontaneous,crepel2023fci,wang2023fractional,reddy2023fractional,2023arXiv230809697X,2023arXiv230914429Y,PhysRevLett.131.136501,PhysRevLett.131.136502,morales2023magic}. The QAH and the FQAH effects in the pentalayer graphene likely arise from a different mechanism, with the  Chern band itself generated from an interaction driven spontaneous crystal formation\cite{dong2023theory,zhou2023fractional,dong2023anomalous,guo2023theory,kwan2023moir}. In this work we restrict to twisted MoTe$_2$ system with narrow Chern band formed at the single particle level, but similar physics may also be possible in interaction driven Chern bands.

More recently, fractional quantum spin Hall(FQSH) effect was reported in twisted MoTe$_2$ system at the hole filling $n=3$ with a smaller twist angle $\theta=2.1^\circ$\cite{kang2023fqsh}. In contrast to the commonly seen valley polarized states at odd integer fillings with $\theta \approx 3.7^\circ$, here the state remains time reversal invariant and supports helical edge modes with the edge conductance $G=\frac{3}{2} \frac{e^2}{h}$. The hole fillings $n=2,4,6$ are consistent with QSH insulators with $G=1,2,3$ in units of $\frac{e^2}{h}$, indicating that the first three moir\'e bands all have the same Chern number $C=1$ in the same valley, while the other valley has $C=-1$ due to the time reversal symmetry. The $n=3$ filling can be naturally understood as the filling $n=\frac{1}{2}+\frac{1}{2}$ of the second moir\'e band on top of the $n=2$ QSH insulator with the first moir\'e band fully filled.  Therefore the experimental result implies a phase with half  quantum spin Hall effect at half filling of the second band in each valley.  Such a phase can not be a Slater determinant state and must be fractionalized. To our best knowledge, this is the first experimental observation of a time reversal invariant fractional phase in the moir\'e system and clearly requires a new theoretical picture beyond the familiar quantum Hall systems.

We can restrict to the $n=\frac{1}{2}+\frac{1}{2}$ filling of the second moir\'e band. Given that the two valleys have Chern number $C=+1, -1$ respectively, the physics is similar to a quantum Hall bilayer\cite{eisenstein2014exciton}, but with opposite chiralities in the two layers. One obvious candidate for FQSH insulator is a decoupled phase with a pair of quantum Hall states in the two valleys. At half filling, we can have a pair of composite Fermi liquids (CFL)\cite{halperin1993theory} or a pair of non-Abelian Pfaffian states\cite{read2000paired}. Both of them have half QSH effect, but the decoupled CFL phase is compressible and does not align with the experimental results indicating a bulk charge gap\cite{kang2023fqsh}. The decoupled Pfaffian phase is consistent with the current experimental data, but faces the following challenges: (1) In the decoupling picture, one should also expect FQSH insulators at fillings like $n=\frac{2}{3}+\frac{2}{3}$. However, the experimental observations of the QSH effect are limited to the integer fillings only. (2) No gapped fractional Chern insulator (FCI) is reported at the filling $n=\frac{5}{2}$. It is not clear why a pair of the gapped FCI can be more stable. (3) The decoupled phases have no inter-valley correlations to reduce the  coulomb repulsion, so they are not energetically favorable compared to competing states such as inter-valley-coherent (IVC) insulator. 

The above observations motivate us to look for an alternative explanation of the experiment. At filling $n=\frac{1}{2}+\frac{1}{2}$, we can start from a pair of composite Fermi liquid (CFL) in the two valleys with opposite chiralities. This conjugate CFL has also been discussed in Ref.~\onlinecite{myerson2023conjugate}. Here we are interested in the possible proximate insulators due to a large inter-valley repulsion. In the conventional quantum Hall bilayer with the same chirality, the ground state is the exciton condensation phase\cite{eisenstein2014exciton}, which can be conveniently described by cooper pairing of the composite fermions\cite{sodemann2017composite}. In contrast, in our case, the exciton feels an effective flux and there is an obstruction to exciton condensation in the flat band limit with weak superlattice effects. Instead, we show that a vortex liquid of the exciton can be  reached by exciton pairing of the composite fermions.  The resulting phase has a charge gap, but hosts neutral Fermi surfaces and gapless spin excitations, resembling a quantum spin liquid in a Mott insulator. 

We dub this new fractional insulator as vortex spin liquid (VSL) because the spin is carried by the internal flux of a U(1) gauge field and the neutral fermions are  vortices of a nearby inter-valley-coherent (IVC) order.  The charge gap here is similar to the Mott gap and suppresses simultaneous occupancy of charges in the two valleys at the same position.  Therefore the phase should be energetically much better in reducing the inter-valley repulsion than the simple decoupled phases.  Meanwhile, the VSL phase still supports half QSH effect and helical charged edge modes in agreement with the experiment\cite{kang2023fqsh}.   We propose future experiments  to test this phase through detecting the separation of the spin and charge gap, metallic spin susceptibility and spin transport, quantum oscillations and thermal Hall effect under Zeeman field.

\textbf{Conjugate composite Fermi liquid} We consider a moir\'e superlattice with valley Chern number $C=1$. For example, in twisted MoTe$_2$ bilayer, the valley K (locked to spin up) and valley K$'$ (locked to spin down) electrons are in a narrow Chern band with Chern numbers $C=1$ and $C=-1$ respectively. At filling $n=\frac{1}{2}+\frac{1}{2}=1$, we restrict to the valley unpolarized states. First, let us turn off the inter-valley repulsion by hand.  Then we have two half-filled Chern bands with opposite chiralities. One obvious phase is a pair of CFLs in the two valleys. We will start from the field theory of this conjugate CFL phase and then add inter-valley correlations.

We describe each CFL by the standard Halperin-Lee-Read (HLR) theory from flux attachment\cite{halperin1993theory}. In the following we label the $K$, $K'$ valley as $+,-$. The HLR theory of the conjugate CFL (cCFL) is:

\begin{align}
    \mathcal L_{\text{cCFL}}&=L_{\text{FS}}[f_\pm,a_\pm]+\frac{1}{8\pi}(a_+da_+-a_-da_-) \notag \\ 
    &~~~-\frac{1}{4\pi}(A_+da_+-A_-da_-)+\frac{1}{8\pi}(A_+dA_+-A_- dA_-)
\end{align}
where $A_{\pm}$ is the abbreviation of the probing U(1) gauge fields $A_{\mu}^\pm$ for the U(1) global symmetry corresponding to the charge conservation of the two valleys respectively. $a_\pm$ is the abbreviation of the emergent dynamical U(1) gauge fields $a_\mu^{\pm}$.   $adb$ is the abbreviation of the Chern-Simons term $\epsilon_{\mu \nu \rho}a_\mu \partial_\nu a_\rho$ with the $\epsilon$ as the anti-symmetric tensor.

The composite fermions $f_\tau$ just form decoupled Fermi surfaces coupled to the internal U(1) gauge fields:

\begin{align}
    \mathcal L_{\text{FS}}[f_\pm,a_\pm]&=\sum_{\tau=\pm} f_\tau^\dagger(t,\mathbf x)(i\partial_t+a_0^\tau+\mu)f_\tau(t,\mathbf x)\notag \\ 
    &~~~+\frac{\hbar^2}{2m} \sum_{\tau=\pm}f^\dagger_\tau(t,\mathbf x)(-i\partial_\mu-a_\mu^\tau)^2f_\tau(t,\mathbf x)
\end{align}
where we use the Minkowski metric. We assume simple $\frac{k^2}{2m}$ dispersion for the fermion and ignore the lattice effect for now. The chemical potential is introduced to fix the density of $f_\tau$ to be half filling per moir\'e unit cell. Within each unit cell there is on-average one flux for each valley.

In the theory, now $f_\tau$ should be viewed as a composite fermion in the valley $
\tau=\pm$. The above action is invariant under the time reversal symmetry: $f_+(t,x) \rightarrow f_-(-t,x)$, $(a_0^+(t,x),\vec a^+(t,x))\rightarrow (a_0^{-}(-t,x),-\vec a^{-}(-t,x)$, $(A_0^+(t,x),\vec A^+(t,x))\rightarrow (A_0^{-}(-t,x),-\vec A^{-}(-t,x))$. It is also useful to redefine $A_\mu^c=\frac{1}{2}(A_\mu^+ + A_\mu^{-})$ and $A_\mu^s=A_\mu^{+}-A_\mu^{-}$. The corresponding charges are $(Q_c, Q_s)=(Q_++Q_-, \frac{Q_+-Q_-}{2} )$. One can see that they are just the charge and the spin $S_z$, while $Q_+,Q_-$ are charges in each valley.  In the following we also use $(Q,S_z)$ to indicate $(Q_c,Q_s)$.

\textbf{VSL from exciton pairing of CFs} The cCFL phase is well justified only in the decoupling limit. Now let us add inter-valley repulsion $H'=\sum_{\mathbf q}V(\mathbf q) \rho_+(\mathbf q)\rho_-(-\mathbf q)$, where $\rho_\tau(\mathbf q)$ is the density operator from the valley $\tau$ and $V(\mathbf q)$ is the coulomb repulsion.  In the conjugate CFL, the density is carried by the internal flux: $\rho_+(\mathbf x)=\frac{1}{4\pi} d  a_+$ where $d  a_+$ is an abbreviation of $\partial_x a_y^+-\partial_y a_x^+$. Similarly $\rho_-(\mathbf x)=-\frac{1}{4\pi} d  a_-$. We  define $a_c=\frac{a_++a_-}{2}$ and $a_s=\frac{a_+-a_-}{2}$.  One then finds that the inter-valley repulsion results in a term proportional to $-\sum_{\mathbf q}V(\mathbf q) q^2 |\mathbf a_c(\mathbf q)|^2+\sum_{\mathbf q}V(\mathbf q)q^2|\mathbf a_s(\mathbf q)|^2$, where $\mathbf a_{c,s}$ here is the transverse component.  As a result, $\mathbf a_s$ is suppressed relative to $\mathbf  a_c$.  It is known that $\mathbf a_s$ and $\mathbf a_c$ mediate attractive and repulsive  between the composite fermions from the two valleys\cite{sodemann2017composite} .  Then in our case the repulsive interaction wins and there is no inter-valley pairing instability of the composite fermions.  This is in contrast to the familiar quantum Hall bilayer with the same chirality in the two layers\cite{sodemann2017composite}.

With inter-valley repulsion comparable to the intra-layer repulsion as in the experiment, there should be an instability of the conjugate CFL phase. Because we are at total filling $n=1$, it is natural to expect a metal to insulator transition when increasing the inter-valley repulsion.  One way to obtain a charge gap is to add intra-valley pairing of the composite fermions, leading to a pair of non-abelian Pfaffian phases\cite{moore1991nonabelions,read2000paired}. However, this decoupled phase does not reduce the inter-valley repulsion.  Instead, we are interested in  searching for an insulator with non-trivial inter-valley correlations.

The cCFL phase is compressible and metallic with the internal fluxes carrying charges $(Q_+,Q_-)=\frac{1}{4\pi}(da_+,-da_-)$.  So one can creates a flux $da_\tau$ to induces a physical charge in the valley $\tau=\pm$.  We also have $(Q,S_z)=\frac{1}{2\pi}(da_s,\frac{1}{2}da_c)$. So internal flux of $d a_s$ induces equally occupied physical charges in the two valleys. Actually $da_s=4\pi$ leads to a cooper pair of electrons.  On the other hand, $da_c=4\pi$ corresponds to a neutral inter-valley exciton with $Q=0,S_z=1$. Due to the inter-layer repulsion, the double occupancy of the two valleys at the same position should be suppressed. It is thus energetically favorable to gap out $a_s$. One natural choice is to consider an exciton pairing of the composite fermions in the mean field Hamiltonian:

\begin{equation}
    H_M'=-\Phi (f^\dagger_+f_-+f^\dagger_- f_+)
\end{equation}

The above term higgses the two internal U(1) gauge fields down to $a_\mu^+=a_\mu^{-}=a^c_\mu$. Meanwhile $a_\mu^s$ is gapped out due to the Meissner effect, leading to a physical charge gap. Now the internal chern-simons  terms are cancelled.   Also, the term $\Phi$ hybridizes the two Fermi surfaces and we can now define $f_R=\frac{1}{\sqrt{2}}(f_++f_-)$ and $f_L=\frac{1}{\sqrt{2}}(f_+-f_-)$. In the following we use $a_\mu=a^c_\mu$ for simplicity. The action of composite fermion is now

\begin{align}
    \mathcal L_{\text{FS}}[f,a]&=\sum_{a=R,L} f_a^\dagger(t,\mathbf x)(i\partial_t+a_0+\mu\pm \Phi)f_a(t,\mathbf x)\notag \\ 
    &~~~-\frac{\hbar^2}{2m} \sum_{a=R,L}f^\dagger_a(t,\mathbf x)(-i\partial_\mu-a_\mu)^2f_a(t,\mathbf x) \notag \\ 
    &~~~+\sum_{a=R,L}\sum_{j=1,...,6}V_\tau e^{i\theta^a_j} f^\dagger_{a}(\mathbf k+\mathbf G)f_a(\mathbf k)
    \label{eq:VSL_dispersion}
\end{align}
where the dispersions from $f_R$ and $f_L$ now have energy shift $-\Phi$ and $+\Phi$ respectively.  We also add a superlattice potential for the composite fermions from the moir\'e lattice. Here $G_1=(\frac{4\pi}{\sqrt{3}a_M},0)$ is a reciprocal vector with $a_M$ the lattice constant. Other $G_j$ with $j=2,3,4,5,6$ can be generated by $C_6$ rotation. Note time reversal now acts as $f_R(\mathbf k)\rightarrow f_R(-\mathbf k), f_L(\mathbf k)\rightarrow -f_L(-\mathbf k)$. $V_a$ and $\theta_j^a$ are the amplitude and phase of the lattice potential for $a=R,L$. We hae $\theta^a_1=\theta^a_3=\theta^a_5=-\theta^a_2=-\theta^a_4=-\theta^a_6=\theta_a$.  $\Phi, V_a, \theta_a$ should be viewed as variational parameters.

\begin{figure}[ht]
    \centering
    \includegraphics[width=0.5\textwidth]{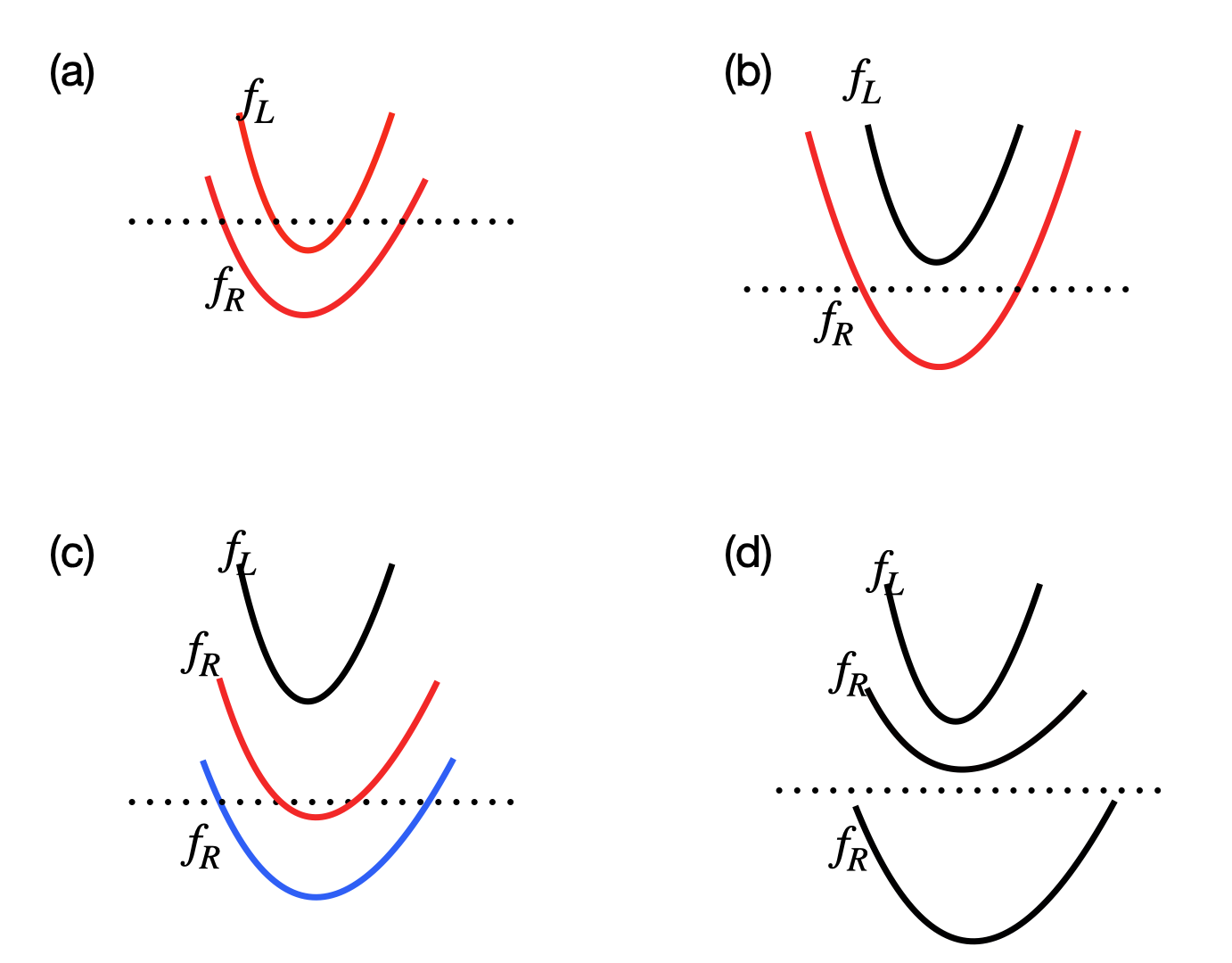}
    \caption{Illustration of the dispersions of the neutral fermion $f_R, f_L$ in Eq.~\ref{eq:VSL_dispersion}. Horizontal dashed line labels the Fermi level. Black, red and blue colors mean that the corresponding bands cross the Fermi level without any Fermi surface, with electron pocket and hole pocket respectively. (a) $\Phi$ and $V_{R/L}$ are small, we have two split Fermi surfaces. (b) $\Phi$ is large and $V_R$ is small. There is one electron pocket from $f_R$ with volume $A_{FS}=1$ in units of the moir\'e Brillouin zone (MBZ. (c) $\Phi$ is large, $V_R$ is intermediate. We have equal size of electron and hole pockets for $f_R$ due to mini bands from the lattice potential. (d) $\Phi$ is large and $V_R $ is large. Now $f_R$ is also gapped due to the band gap opening from the lattice effect.  }
    \label{fig:VSL_dispersion}
\end{figure}

The total Fermi surface volumes should be $1$ in units of the moir\'e brillouin zone (MBZ) for $f_R, f_L$ together. When the lattice effect is not large, such as in the perfect Landau level, even a large $\Phi$ can only gap out $f_L$ and leave a single Fermi surface for $f_L$ (see Fig.~\ref{fig:VSL_dispersion}(a)(b)). When there is large variation of effective magnetic field inside the moir\'e unit cell, the composite fermions $f_R, f_L$ feel a superlattice potential $V_R, V_L$. When we increase $V_R$, initially the Fermi surface in $f_R$ can not be gapped and we expect a compensated semi-metal with equal size of electron and hole pockets as shown in Fig.~\ref{fig:VSL_dispersion}(c). Note that the total Fermi surface volume is $1$ or $0$ in units of the moir\'e Brillouin zone.  When the lattice effect is large enough, $f_R$ is also fully gapped as in Fig.~\ref{fig:VSL_dispersion}(d). The situations in Fig.~\ref{fig:VSL_dispersion}(a)(b)(c) give roughly the same physics and we unify them within the same phase without distinguishing their detailed differences in dispersions and Fermi surfaces. When the fermions are fully gapped as in Fig.~\ref{fig:VSL_dispersion}(d), we have a different phase, a superfluid phase of excitons (or equivalently an inter-valley coherent insulator) which we discuss later.

The final theory is

\begin{align}
\mathcal L_{\text{VSL}}=\mathcal L_{FS}[f,a]-\frac{1}{4\pi}A_s d a+\frac{1}{4\pi}A_c d A_s
\label{eq:VSL}
\end{align}
where we have used the definition $A_c=\frac{1}{2}(A_++A_-)$ and $A_s=A_+-A_-$.   One can also start from the Dirac theory of the CFL\cite{son2015composite,wang2016half,metlitski2016particle,wang2015dual} and consider a similar phase from exciton pairing of the composite fermions. In the supplementary we show that essentially the same theory will be reached following the Dirac theory.

 The theory in Eq.~\ref{eq:VSL} describes a fractional insulator with gapless spin degree of freedom (or neutral excitons).  First, let us ignore the background term $\frac{1}{4\pi} A_c d A_s$.  Then the remaining action couples only to $A_s$ and it should be viewed as a quantum spin liquid. In this spin liquid, there are gapless spin excitations, but they are carried by the internal flux: $da=2\pi$ flux carries spin $S_z=-\frac{1}{2}$. The double monopole operator $\mathcal M_a^2$ of the gauge field $a_\mu$ now corresponds to the physical exciton creation operator with $S_z=1$.  On the other hand,  the fermions $f_R, f_L$ are neutral and spinless. They should be viewed as fermionic vortices of a nearby exciton condensation phase, an inter-valley-coherence (IVC) order in the context of the twisted MoTe$_2$ system. For this reason we dub this spin liquid as vortex spin liquid (VSL).  It is in certain sense a composite fermi liquid (CFL) of the physical excitons. For excitons, the physical time reversal symmetry acts as an anti-unitary particle-hole symmetry and forbids the internal Chern-Simons term, similar to Son's Dirac theory of the CFL in the half-filled Landau level with particle-hole symmetry\cite{son2015composite}.

\textbf{Bulk property of the VSL phase} Similar to the familiar U(1) spin liquid with spinon fermi surface, the VSL phase is metallic in thermal transport and has a finite spin susceptibility $\chi_s$. Unlike the spinon Fermi surface state, only  the correlation function of $S_z$, $\langle S_z(\mathbf x)S_z(0)\rangle \sim \frac{1}{|\mathbf x|^\alpha}$ has a finite exponent $\alpha$.  On the other hand, $\langle S^\dagger(\mathbf x) S^{-}(0)\rangle$ has an infinite power-law exponent because $S^\dagger$ corresponds to the monopole operator of the gauge field $a_\mu$, which has infinite scaling dimension due to coupling to the Fermi surface\cite{lee2008stability}. Another unique property of the VSL phase is that its spin transport resistivity $\rho_s=\frac{4}{\rho_f}$ (see the supplementary) is inverse to  $\rho_f$, the resistivity of the neutral fermions.  As we expect $\rho_f$ increases with the temperature $T$ as in a usual metal, the spin resistivity $\rho_s$ increases when decreasing the temperature until it saturates at a value $\frac{4}{\rho_f(T=0)}$ at zero temperature.  Actually if there is no disorder, the residual resistivity $\rho_f(T=0)=0$ and then $\rho_s$ is infinite at zero temperature like an insulator. In this sense, disorder can decreases the resistivity of the spin (exciton) transport and enhances the spin (exciton)  mobility. This is because the neutral fermion $f$ is vortex of a nearby superfluid of the excitons and its movement destroys the superfluidity of the excitons. Disorder helps localizing these vortices and reduces the resistivity of the spin (exciton) transport, similar to the type II superconductor. 

Another interesting property of the VSL phase is that it generates an internal flux $da$ under zeeman field. Because of the finite uniform spin susceptibility $\chi_s$, we have $S_z=g \chi_s B$ with $B$ the Zeeman field and $g$ the renormalized g-factor. In the VSL phase the internal flux $b=da=\partial_x a_y-\partial_y a_x= 4\pi S_z=4\pi g \chi_s B$.  Therefore the neutral fermions feels an effective magnetic field $b_{eff}=\alpha B$ with $\alpha=4\pi g \chi_s$ and thus there should be quantum oscillations in this insulator.  Note that $\alpha$ in principle can be much larger than $1$.  If $g$ factor is large and $\chi_s \sim m_f$ is large, we can imagining that a small magnetic field (at order of 1 Tesla) can almost fully polarize $S_z$ given that the energy scale in the spin sector can be small. This will give one effective flux per moir\'e unit cell for the neutral fermions, which normally needs around 30 Tesla with the orbital effect of the magnetic field even for moir\'le lattice constant $a_M\sim 15$ nm.   From this logic, we may expect large effective flux at small external magnetic field and quantum oscillations under the Zeeman field.  For the same reason, there should be a finite thermal Hall conductivity $\kappa_{xy}$ proportional to the Zeeman field.  Our mechanism of generating internal flux is  through the Zeeman effect and thus can work in the infinite charge gap limit. This is different from the orbital-magnetic effect for spinon Fermi surface state\cite{motrunich2006orbital} which only works in the limit of small charge gap. The major worry to observe these effects is that the Zeeman field may quickly favor the valley polarized phase through a first order transition. 

\textbf{FQSH effect and helical edge mode} The VSL phase here in Eq.~\ref{eq:VSL} is similar to a Mott insulator with gapless spin liquid, but it has one unusual property beyond the conventional Mott insulator with well defined localized Wannier orbital. Due to the background term $\frac{1}{4\pi}A_c d A_s$, there is a fractional quantum spin Hall effect. Note here the coefficient is half of the QSH insulator at $n=2$: a flux $dA_c=2\pi$ creates $S_z=\frac{1}{2}$ in our case. One immediate consequence is that there must be helical edge mode carrying charge $Q_c$. To see this, we can follow the Laughlin argument on an annulus. By threading a flux $dA_s=4\pi$, a total charge $Q_c=1$ must be transported from the inner edge to the outer edge. Given that the bulk has a charge gap, the flux threading process can be done adiabatically in the charge sector, meaning no gapped charge excitations are created in the bulk. Then the edge must host gapless charge mode to make the transportation of the charge possible. And due to the time reversal symmetry, there must be helical edge mode as in the integer QSH insulator. 

 In the supplementary we show that the edge theory is the same as that from a pair of $\nu=\frac{1}{2}$ bosonic Laughlin state and its time reversal partner:

\begin{align}
    \mathcal S_{\text{edge}}&=\int dx dt \frac{2}{4\pi}  \partial_x \phi_+ \partial_t \phi_{+}-\frac{2}{4\pi} \partial_x \phi_{-} \partial_t \phi_{-}\notag \\ 
    &~~~-\frac{2}{4\pi}\upsilon (\partial_x \phi_+)^2-\frac{2}{4\pi}\upsilon(\partial_x \phi_-)^2-\frac{g}{\pi} \partial_x \phi_+ \partial_x \phi_{-}
    \label{eq:edge_theory}
\end{align}
where $\phi_{\pm}$ are two counter-propagating modes from the valley $\tau=\pm$ respectively. All of the back-scattering terms such as $\cos (\phi_+\pm \phi_-)$ are forbidden by the charge and $S_z$ conservation. There are couplings to the bulk gapless spin degree of freedom. In the supplementary we argue that these couples are irrelevant. For example, we can have a term $S^{(1)}_{\text{edge-bulk}}=-\lambda_1 \int dt dx \mathcal M_a^\dagger e^{i(\phi_++\phi_-)}+h.c.$ which is a combination of the backscattering and a monopole operator creating a $S_z=-\frac{1}{2}$ excitation in the bulk. However, the monopople operator is suppressed by the neutral fermi surface\cite{lee2008stability}, so the term is irrelevant. We also show that the coupling to the internal flux $\partial_x a_y -\partial_y a_x$ through $S^{\text{edge}}_z S^{\text{bulk}}_z$ coupling is irrelevant because the Landau damping term $\frac{|\omega|}{|q_x|}$ in the propagator of the gauge field $a_\mu$ makes $a_\mu$ effectively massive for the edge with linear dispersion $\omega \sim q_x$ (see the supplementary). We  thus conclude that the edge mode decouples from the gapless bulk. 

\textbf{Connection to IVC insulator} At filling $n=\frac{1}{2}+\frac{1}{2}$, if we restrict to Slater determinant states, the most obvious time reversal invariant insulator is an excitonic insulator with inter-valley-coherent (IVC) order. However, the inter-valley exciton feels an effective two flux per moir\'e unit cell and its condensation is frustrated. If the superlattice effects are strong, we can imagine a vortex crystal state of the exciton condensation. But at smaller variation of the effective magnetic field and lattice effects, the vortex crystal may be melted to a vortex liquid phase. When the vortices move, the superfluidity of the excitons is destroyed. The VSL phase we constructed is exactly such a vortex liquid of the excitons. It can be smoothly connected to an IVC insulator by gapping out the Fermi surfaces through increasing the lattice potential $V_R$ as illustrated in Fig.~\ref{fig:VSL_dispersion}(d).  IVC insulator has been reported in Hartree-Fock study\cite{wang2023topology}, the vortex liquid phase proposed here may be a natural competing state beyond Hartree Fock in the same parameter regime. Even if the IVC state wins due to the large lattice effects, our approach from gapping out neutral fermions may still give an energetically better variational wavefunction than the Hatree Fock wavefunction. We leave to future to explore this possibility.

\textbf{Summary} In conclusion, we propose a new time reversal invariant fractional insulator, dubbed as vortex spin liquid (VSL), at odd integer filling of moir\'e Chern band with Chern number $C=\pm 1$ in each valley.  The VSL phase has a charge gap, but hosts gapless spin excitation, resembling the usual Mott insulator. Interestingly it also supports well-isolated helical charged edge modes and a FQSH effect. We propose the VSL phase as an explanation of  the recent observed FQSH state at filling $n=3$ of twisted MoTe$_2$ at twist angle around $2.1^\circ$\cite{kang2023fqsh}.   As a spin liquid, the physics discussed here can exist only at odd integer filling, also consistent with the experiment. Future experimental measurements of the spin and charge gap separately can potentially distinguish the VSL phase and the simple decoupled FQSH phase in this system. From a theoretical perspective, this new phase offers an example which combines physics of Mott insulator and quantum spin Hall effect, similar to a previous proposal of quantum Hall spin liquid\cite{zhang2020quantum}, which combines Mott physics and quantum Hall effect. A systematic understanding of possible quantum spin liquid states enriched by topological charge response at edge should be a promising future research direction.

\textit{Note added:} The current manuscript is a new version of an unpublished preprint in 2018\cite{zhang2018composite}. The VSL phase here was called composite fermion insulator in the previous version, but the essential theory remains the same.

\textbf{Acknowledgement} This work was supported by the National Science Foundation under Grant No. DMR-2237031.

\bibliographystyle{apsrev4-1}
\bibliography{vsl}

\onecolumngrid
\appendix

\section{Derivation of the HLR theory of the conjugate CFL}

It is more convenient to derive the HLR theory using the slave boson construction.   For each valley $\tau=+,-$, we do parton construction $c_\tau(\mathbf x)=b_\tau(\mathbf x) f_\tau(\mathbf x)$, where $c_\tau$ is the electron operator. $b_\tau$ and $f_\tau$ are the slave boson and the neutral fermion operator. There are two U(1) gauge fields $a_\mu^{+},a_\mu^{-}$ from the gauge transformation  $b_\tau(\mathbf x) \rightarrow b_\tau(\mathbf x) e^{-i \theta_\tau(\mathbf x)}, f_\tau(\mathbf x) \rightarrow f_\tau(\mathbf x) e^{i \theta_\tau(\mathbf x)}$ which leaves the electron operator invariant. Meanwhile there are two independent global U(1) symmetry: $c_\tau(\mathbf x)\rightarrow c_\tau(\mathbf x)e^{i\theta_\tau}$. We assign the charge in the following way: $b_\tau(\mathbf x)\rightarrow b_\tau(\mathbf x)e^{i\theta_\tau},f_\tau(\mathbf x)\rightarrow f_\tau(\mathbf x)$ so $f_\tau$ is neutral under both U(1) global symmetry.  We introduce two U(1) probing field $A_\mu^{\pm}$ for the two U(1) symmetry respectively.  In summary, $b_\tau$ couples to $-a_\mu^{\tau}+A_\mu^{\tau}$ while $f_\tau$ couples to $a_\mu^{\tau}$. It is also useful to redefine $A_\mu^c=\frac{1}{2}(A_\mu^+ + A_\mu^{-})$ and $A_\mu^s=A_\mu^{+}-A_\mu^{-}$. The corresponding charges are $(Q_c, Q_s)=(Q_++Q_-, \frac{Q_+-Q_-}{2} )$. One can see that they are just the charge and the spin $S_z$, while $Q_+,Q_-$ are charges in each valley.  The system also has a time reversal symmetry: $c_+(\mathbf x)\leftrightarrow c_-(\mathbf x), b_+(\mathbf x)\leftrightarrow b_-(\mathbf x), f_+(\mathbf x) \leftrightarrow f_-(\mathbf x)$.

 The low energy theory of the conjugate CFL (cCFL) is:
\begin{align}
\mathcal L_{\text{cCFL}}=\mathcal L_b[b,A-a]+\mathcal L_{\text{FS}}[f_\pm,a_\pm]
\label{eq:decouple_CFLs_appendix}
\end{align}

Here $f_\tau$ just form decoupled Fermi surfaces:

\begin{align}
    \mathcal L_{\text{FS}}[f_\pm,a_\pm]=\sum_{\tau=\pm} f_\tau^\dagger(t,\mathbf x)(i\partial_t+a_0^\tau+\mu)f_\tau(t,\mathbf x)
     -\frac{\hbar^2}{2m} \sum_{\tau=\pm}f^\dagger_\tau(t,\mathbf x)(-i\partial_\mu-a_\mu^\tau)^2f_\tau(t,\mathbf x)
\end{align}
where we have ignored the lattice effect for now.

The slave boson $b_\tau$ should be put in a pair of Laughlin state:

\begin{align}
    \mathcal L_b[b,A-a]=-\frac{2}{4\pi}\alpha_+ d \alpha_++\frac{1}{2\pi} (A_+-a_+)d\alpha_+ 
    +\frac{2}{4\pi}\alpha_- d \alpha_-+\frac{1}{2\pi} (A_{-}-a_{-})d\alpha_{-}
    \label{eq:laughlin}
\end{align}
where $a_{\pm}$ is an abbreviation of $a^{\pm}_\mu$, another two U(1) gauge fields to capture the topological order of the Laughlin states. $adb$ is the abbreviation of the Chern-Simons term $\epsilon_{\mu \nu \rho}a_\mu \partial_\nu a_\rho$ with the $\epsilon$ the anti-symmetric tensor.

We can integrate $\alpha_\pm$ and reach the HLR theory of the conjugate CFL (cCFL):

\begin{align}
    \mathcal L_{\text{cCFL}}=\mathcal L_{\text{FS}}[f_\pm,a_\pm]+\frac{1}{8\pi}(a_+da_+-a_-da_-) -\frac{1}{4\pi}(A_+da_+-A_-da_-)+\frac{1}{8\pi}(A_+dA_+-A_- dA_-)
    \label{eq:ccfl_full_theory_appendix}
\end{align}

\section{Vortex spin liquid from Dirac theory of CFL}

Recently a Dirac theory has been proposed to describe the CFL phase in the half-filled Landau level based on a fermionic particle-vortex duality\cite{son2015composite,wang2016half,metlitski2016particle,wang2015dual}. In the Dirac picture, the effective theory for the two decoupled CFLs with opposite chirality is: 

\begin{align}
  \mathcal L_{\text{cCFL}}=\sum_{\tau=\pm}\left(\bar \Psi_\tau(i\slashed \partial+\slashed a_\tau)\Psi_\tau+\frac{1}{4\pi} A_\tau d a_\tau\right)
  -\mu \Psi^\dagger \tau_z \Psi   +\frac{1}{8\pi}A_+ d A_+-\frac{1}{8\pi}A_- d A_-
\end{align}
where $\tau=\pm$ is the layer index. $\tau_\mu$ is the Pauli matrix in the valley space.

In the Dirac picture, fermions $\Psi=(\Psi_+,\Psi_-)$ are vortices and their densities are fixed by the external magnetic flux. For our case with opposite chirality, we have a electron pocket in one valley and a hole pocket in the other valley.  This is why the chemical potential is added as a $-\mu \tau_z$ term. In this theory, the time reversal acts as: $\Psi(t,\mathbf{x}) \rightarrow \tau_x \sigma_y \Psi^\dagger(-t,\mathbf{x})$, $a^0(t,\mathbf{x})\rightarrow -\tau_x a^0(-t,\mathbf{x})$, $\vec{a}(t,\mathbf{x})\rightarrow \tau_x \vec{a}(-t,\mathbf{x})$. Here $\sigma_a$ acts in the Dirac spinor space. The above action also has a particle-hole symmetry which acts as $\Psi(t,\mathbf{x}) \rightarrow i\sigma_y \Psi(-t,\mathbf{x})$, $a^0(t,\mathbf{x})\rightarrow a^0(-t,\mathbf{x})$, $\vec{a}(t,\mathbf{x})\rightarrow -\vec{a}(-t,\mathbf{x})$. However, with the lattice effects, we expect the particle-hole symmetry to be broken. So we will ignore the particle-hole symmetry in the following discussions.

For convenience, we make a  redefinition: $\Psi_- \rightarrow \Psi_-^\dagger$ and $a^{-}_\mu \rightarrow -a^{-}_\mu$,  then the final effective theory is:
\begin{align}
  \mathcal L_{\text{cCFL}}=\bar \Psi_+(i\slashed \partial+\slashed a_+)\Psi_+-\mu \Psi^\dagger_+ \Psi_++\frac{1}{4\pi} A_+ d a_+
  +\bar \Psi_{-}(i\slashed \partial+\slashed a_{-})\Psi_{-}-\mu \Psi^\dagger_{-} \Psi_{-}-\frac{1}{4\pi} A_{-} d a_{-}
  +\frac{1}{8\pi}A_+ d A_+-\frac{1}{8\pi}A_{-} d A_{-}
  \label{eq:dirac_theory_full}
\end{align}

We will use Eq.~\ref{eq:dirac_theory_full} as our starting point. We still define $\Psi=(\Psi_+,\Psi_-)$ with the new definition of $\Psi_2$. Now the  time reversal acts as: $\Psi(t,\mathbf{x}) \rightarrow \tau_x \sigma_y \Psi(-t,\mathbf{x})$, $a^0(t,\mathbf{x})\rightarrow \tau_x a^0(-t,\mathbf{x})$, $\vec{a}(t,\mathbf{x})\rightarrow -\tau_x \vec{a}(-t,\mathbf{x})$.

In this new representation, we have two electron Fermi pockets from the two valleys. We still define $a_c=\frac{a_1+a_2}{2}$ and $a_s=\frac{a_1-a_2}{2}$. $A_c$ and $A_s$ follow the same convention as in the main text. Then $d a_s$ carries the physical charge under $A_c$.  Inter-valley repulsion suppresses $d a_s$ and the fluctuation of $a_c$ dominates.  However, in our case $\Psi_+$ and $\Psi_-$ carry the same charge under $a_c$. As a result, $a_c$ fluctuation mediates repulsive interaction and suppresses the inter-valley pairing instability of the composite fermions. We reach the conclusion that the conjugate CFL phase described by the Dirac theory in Eq.~\ref{eq:dirac_theory_full} is stable against a small inter-layer repulsion.

At larger inter-valley repulsion, we expect a charge gap opening from gapping out $a_s$. This can be done by the exciton condensation between the Dirac composite fermions: $\langle \Psi_{+}^\dagger(\mathbf k) A(\mathbf k) \Psi_{-} (\mathbf k) \rangle \neq 0$, where $A(\mathbf k)$ is a $2\times 2$ matrix depending on the momentum $\mathbf{k}$.  Here we propose the simplest term $-\Phi \Psi^\dagger_{+}(\mathbf k) \Psi_{-}(\mathbf k)\rangle$. One can check that it is invariant under both time reversal and the particle-hole transformation (if the particle-hole symmetry exists). The condensation higgses $a_+=a_{-}=a$. The final theory is:

\begin{align}
  \mathcal L_{\text{VSL}}&=\sum_{\tau=\pm}\left(\bar \Psi_\tau(i\slashed \partial-\slashed a)\Psi_\tau-\mu \Psi^\dagger_\tau \Psi_\tau \right)-(\Phi \Psi_1^\dagger \Psi_2+h.c.)-\frac{1}{4\pi} A_s d a+\frac{1}{4\pi}A_c d A_s
  \label{eq:dirac_theory_CFI}
\end{align}

This is very similar to the action for the vortex spin liquid phase in the main text.  The term $\Phi$ still hybridizes the two Fermi surfaces to split them. Because of the time reversal symmetry, the Fermi surface after hybridization does not have any berry phase. So it does not matter whether the original dispersion of the composite Fermions has a $\pi$ berry phase or not.  One can also add mass term $M \bar \Psi \tau_z \Psi$ to break the partile-hole symmetry and give a mass for the Dirac fermions.  But in the final theory of the VSL phase, the Berry curvature is cancelled due to the hybridization $\Phi$. We conclude that the Dirac theory of the CFL does not give any difference in describing the VSL phase from the approach in the main text. We will use HLR theory in the rest of the paper.

\section{Fractional quantum spin Hall effect and edge theory of the VSL phase}

To better characterize the edge mode and FQSH of the VSL phase, we start from the full action in Eq.~\ref{eq:decouple_CFLs_appendix}. We expect helical edge mode from the slave boson theory $\mathcal L_b[b,A-a]$. We make a redefinition $\alpha_\pm \rightarrow \tilde \alpha_\pm \mp \frac{1}{2} a_\pm$ to reach a new form of the theory:

\begin{align}
    \mathcal L_{\text{cCFL}}&=\mathcal L_{\text{FS}}[f_\pm,a_\pm]+\frac{1}{8\pi}(a_+da_+-a_-da_-) -\frac{1}{4\pi}(A_+da_+-A_-da_-) +\mathcal L_{\text{FQSH}}\notag \\
   \mathcal L_{\text{FQSH}} &=-\frac{2}{4\pi}\tilde \alpha_{+} d \tilde \alpha_{+}+\frac{2}{4\pi}\tilde \alpha_{-} d \tilde \alpha_{-}+\frac{1}{2\pi}A_+d\tilde \alpha_++\frac{1}{2\pi} A_{-} d \tilde \alpha_{-}
\end{align}

If we integrate $\tilde \alpha_\pm$, we will generate the background term $\frac{1}{8\pi}(A_+ d A_+-A_{-} d A_{-})=\frac{1}{4\pi} A_c d A_s$ and recover Eq.~\ref{eq:ccfl_full_theory_appendix}. One can see that the role of $\tilde \alpha$ is to provide the background fractional quantum spin Hall effect. Here we will keep them as they generate the edge modes.

Now after the exciton condensation of the composite fermions, we lock $a_+=a_-=a$ and the theory of the VSL phase is:

\begin{align}
    \mathcal L_{\text{VSL}}&=\mathcal L_{\text{FS}}[f,a]-\frac{1}{4\pi}A_s d a +\mathcal L_{\text{FQSH}}\notag \\
   \mathcal L_{\text{FQSH}} &=-\frac{2}{4\pi}\tilde \alpha_{+} d \tilde \alpha_{+}+\frac{2}{4\pi}\tilde \alpha_{-} d \tilde \alpha_{-}+\frac{1}{2\pi}A_+d\tilde \alpha_++\frac{1}{2\pi} A_{-} d \tilde \alpha_{-}
\end{align}

Note that the time reversal symmetry act as: $a_0(t,\vec x)\rightarrow a_0(-t,\vec x), \vec a(t,\vec x) \rightarrow -\vec a(t,\vec x)$, $\alpha^+_0(t,\vec x) \rightarrow  -\alpha^{-}_0(-t,\vec x), \vec \alpha_+ (t,\vec x) \rightarrow \vec \alpha_-(t,\vec x)$, $A^s_0(t,\vec x) \rightarrow -A^s_0(-t,\vec x), \vec A_s(t,\vec x) \rightarrow \vec A_s(-t,\vec x)$.


The part of $\mathcal L_{\text{FQSH}}$ gives the standard helical edge theory. Suppose the system is at $y>0$, then the edge theory at $y=0$ is described by the following action:

\begin{equation}
    \mathcal S_{\text{edge}}=\int dx dt \frac{2}{4\pi}  \partial_x \phi_+ \partial_t \phi_{+}-\frac{2}{4\pi} \partial_x \phi_{-} \partial_t \phi_{-}-\frac{2}{4\pi}\upsilon (\partial_x \phi_+)^2-\frac{2}{4\pi}\upsilon(\partial_x \phi_-)^2-\frac{2}{4\pi}2g \partial_x \phi_+ \partial_x \phi_{-}
    \label{eq:edge_theory_appendix}
\end{equation}

In the edge we can identify $\alpha^{\pm }_x=\partial_x \phi_{\pm}$.  So the time reversal symmetry acts as $\phi_+ \rightarrow \phi_-$. $\rho_{\pm}=\frac{1}{2\pi}\partial_{x}\phi_{\pm}$ gives the density operator in the valley $\pm$.

We have the commutation relationship:

\begin{align}
    [\phi_{\pm}(x),\partial_{x'}\phi_{\pm}(x')]=\pm\frac{2\pi}{2} \delta(x-x')
\end{align}

As a result, the operator $e^{ \pm i\phi_\pm}$ creates a 1/2 charge for the valley $\pm$ respectively.

Next we define $\varphi=\frac{1}{2}(\phi_++\phi_-)$ and $\theta=\frac{1}{2}(\phi_+-\phi_-)$, then the total charge density is $\rho_c=\frac{1}{\pi}\partial_x \varphi$ and the spin $S_z$ density is $\rho_s=\frac{1}{\pi}\partial_x \theta$.  Now the action becomes

\begin{align}
    S_{\text{edge}}=\int dt dx \frac{2}{\pi}\partial_x \varphi \partial_t \theta-\frac{1}{\pi}\upsilon (1-g)(\partial_x \theta)^2-\frac{1}{\pi}\upsilon(1+g) (\partial_x \varphi)^2
\end{align}

We can also integrate $\varphi$ and reach

\begin{align}
    S_{\text{edge}}=\int dt dx \frac{K}{2\pi \tilde \upsilon} \big( (\partial_t\theta)^2-\tilde \upsilon^2 (\partial_x \theta)^2\big)
\end{align}
with $K=2 \sqrt{\frac{1-g}{1+g}}$ and $\tilde \upsilon=\upsilon \sqrt{1-g^2}$.

In this convention, $\psi_+^\dagger=e^{i(\theta+\varphi)}$ creates a right-moving $1/2$ charge in the valley $+$. $\psi_-^\dagger=e^{i(\theta-\varphi)}$ creates a left-moving $1/2$ charge in the valley $-$.  $e^{2i\varphi}$ creates a $S_z=\frac{1}{2}$ neutral exciton with scaling dimension $K$.  $e^{2i\theta}$ creates a spinless $Q=1$ excitation (half of the physical Cooper pair)
 with scaling dimension $\frac{1}{K}$. Time reversal acts $\theta \rightarrow -\theta$.

 Because $\partial_\mu \theta$ couples to the external probing field $A_c$ as $(\partial_\mu-i \frac{1}{2} A_{c;\mu})\theta$, the conductance is $G=K (\frac{1}{2})^2=\frac{K}{4}=\frac{1}{2} \sqrt{\frac{1-g}{1+g}}$. One interesting question is whether a finite $g$ alters the conductance or not. The question exists even for the integer quantum spin Hall insulator, where the Luttinger parameter can also be modified by the inter-spin interaction.
It turns out that the existence of the lead attached to the system make the conductance unchanged by the interaction $g$\cite{maslov1995landauer}.  We thus expect that the conductance is still $G=\frac{1}{2} \frac{e^2}{h}$ even at finite $g$ in our case, but a detailed modeling of the lead is needed to confirm this conclusion.

\subsection{Irrelevance of the edge-bulk coupling}

Because our bulk has gapless spin excitations, we may worry that the couplings between the edge and the bulk change the edge physics. Here we show that the edge-bulk couplings are irrelevant. The coupling can be grouped to two categories: (I) $S^\dagger S^{-}$ coupling and (II) $S_z S_z$ coupling.

The first type of edge-bulk interaction is

\begin{equation}
    S_{\text{edge-bulk}}^{(1)}=\lambda_1 \int dt dx  e^{2i\varphi} \mathcal M_a^\dagger+h.c.
\end{equation}
where $\mathcal M_a^\dagger$ creates a monopole $da=2\pi$ of the bulk U(1) gauge field $a_\mu$ and carries $S_z=-\frac{1}{2}$. $e^{2i\varphi}$ is  the edge operator which creates an excitation with $S_z=\frac{1}{2}$. Given that the scaling dimension of $e^{2i\varphi}$ is $K=2\sqrt{\frac{1-g}{1+g}}$, the scaling dimension of $\lambda_1$ is negative if $[\mathcal M_a^\dagger]>2(1-\sqrt{\frac{1-g}{1+g}})$.  Because of the neutral Fermi surface, in the bulk the scaling dimension of the monopole operaotr is infinite\cite{lee2008stability}. At the edge the scaling dimension should also be large and positive. We believe $\lambda_1$ term is irrelevant.

The second type of edge-bulk coupling is from the $S_z^{\text{bulk}} S_z^{\text{edge}}$ coupling with $S_z^{\text{bulk}}\sim \partial_x a_y-\partial_y a_x$ in the bulk. This leads to

\begin{equation}
    S^{(2)}_{\text{edge-bulk}}=\lambda_2\int dt dx \partial_x \theta \partial_x a_y
\end{equation}

We know the propagator of $a_y$ in the bulk is $D(\omega,q_x)=\frac{1}{\chi_d q_x^2+\kappa \frac{|\omega|}{|q_x|}}$, where the second term is due to the Landau damping from the neutral Fermi surface. Now in the perspective of the edge with linear dispersion, we should consider the $\omega=\upsilon q_x$ regime and then $\kappa \frac{|\omega|}{|q_x|}\sim \frac{\kappa}{\upsilon}$ is a mass term for the gauge field. As a result, the gauge field behaves massive from the eyes of the edge and $\lambda_2$ term is irrelevant. Integration of  the gauge field $a$ only leads to a term at order $q_x^4 |\theta(\omega,q_x)|^2$.

We have now shown that the spin-spin couplings to the bulk  do not induce any non-trivial interactions in the edge and can be ignored.  Because there is no gapless charge excitation in  the bulk, we believe the bulk does not alter the behavior of the density-density interaction in  the edge. In summary, we conclude that the edge theory in Eq.~\ref{eq:edge_theory_appendix} decouples from the gapless bulk. Therefore, in terms of the charge transport meausrements done in the experiment\cite{kang2023fqsh}, we expect exactly the same behavior as a simple FQSH insulator with just a pair of Luaghlin state and its time reversal partner.

\section{Spin susceptibility and spin transport of the VSL phase}
In this section we discuss the susceptibility and transport of the spin degree of freedom in the vortex quantum spin liquid.  There is a mapping between the spin degree of freedom and the inter-valley exciton: $S^\dagger \leftrightarrow b_e^\dagger, S^{-} \leftrightarrow b_e, S_z \leftrightarrow n_b-\frac{1}{2}$, where $b_e^\dagger$ is the creation operator of the exciton and $n_b$ is the density operator of the exciton.  Due to the mapping, the spin susceptibility and spin transport correspond to the compressibility and transport of the inter-valley exciton.

We consider the action:

\begin{align}
\mathcal L_{VSL}=\mathcal L_{FS}[f,a]-\frac{1}{4\pi}A_s d a+\frac{1}{4\pi}A_c d A_s
\end{align}

$A_s$ is the probing field of the exciton. For now we set $A_c=0$ and can ignore the last background term $A_c d A_s$.  The remaining action resembles a composite Fermi liquid (CFL) of the exciton, but without the Chern-Simons term.  Note that the time reversal symmetry acts as a particle-hole symmetry for the exciton: $b_e \rightarrow b_e^\dagger$, so the above action does not have self chern-simons term and has similar structure as Son's Dirac theory of the CFL at half filled Landau level\cite{son2015composite}, though now the neutral fermion $f$ forms regular parabolic dispersion instead of Dirac dispersion.

In this theory the neutral fermion $f$ is the vortex of the exciton, while the exciton creation operator corresponds to the monopole of the internal U(1) gauge field $a_\mu$, similar to Son's Dirac theory\cite{son2015composite}. The shapes of the neutral Fermi surfaces depend on details such as the superlattice potential the neutral fermions feel. We will focus on universal property independent of the details.

To get the linear response to the external field $A_s$, we first integrate $f$ and obtain an effective action for the internal U(1) field $a_\mu$.  $a_0$ is screened by the finite density of states of the Fermi surface.  The dominant term is the transverse $\mathbf a_\perp(\omega,\mathbf q)$ with $\mathbf q \cdot \mathbf a_\perp=0$.  Then the effective action at the $|\mathbf q|\rightarrow 0$ limit is

\begin{align}
    \mathcal L_{\text{eff}}[A_s,a]&= \frac{1}{2} \sum_{\omega,}\sum_{\mathbf q}(\chi_d |\mathbf q|^2+ \frac{1}{2\pi} i \omega \sigma_{f}) |\mathbf a_\perp(\omega,\mathbf q)|^2+\frac{1}{4\pi} \sum_{\omega,\mathbf q} A_x(\mathbf q)( i\omega)a_y(-\mathbf q)-\frac{1}{4\pi}A_y(\mathbf q)(i\omega) a_x(-\mathbf q)\notag \\ 
    &~~~-\frac{1}{4\pi}\sum_{\omega,q}A_0(\mathbf q)(-iq_x)a_y(-q)-\frac{1}{4\pi}\sum_{\omega,q}A_0(\mathbf q)(iq_x)a_x(-q)
\end{align}
where $\chi_d$ is the diamagnetism from the composite fermions. We should have $\chi_d=\sum_{I} \frac{1}{24\pi m_f^I}$ with $I$ summed over all Fermi surfaces. $\sigma_{f}$ is the conductivity of the neutral fermion.  We have used the natural unit $e=\hbar=1$. Due to the time reversal symmetry, the Hall conductivity of the neutral fermions vanishes. Then integration of $a$ leads to the response action of $A_s$:

\begin{equation}
    \mathcal L_{\text{eff}}[A_s]=\frac{1}{2} \sum_{\omega,\mathbf q} \frac{1}{16\pi^2 \chi_d} |A_0(\mathbf q)|^2+\frac{1}{2} \sum_{\omega,\mathbf q} (\frac{1}{2\pi} i\omega \frac{1}{4 \sigma_f})|\mathbf A_{\perp}(\omega,\mathbf q)|^2
\end{equation}

From this we can read out the exciton compressibility (or spin susceptibility) $\kappa_s=\frac{1}{16\pi^2\chi_d}$ and the spin conductivity $\sigma_s=\frac{1}{4\sigma_f}$ in units of $\frac{e^2}{h}$.  Interesting the spin resistivity $\rho_s=\frac{4}{\rho_f}$ is inverse to the resistivity of the neutral fermions $\rho_f=\frac{1}{\sigma_f}$.   In a clean system, $\rho_f$ is small and thus the spin resistivity $\rho_s$ is large.  For this reason, disorder can actually reduces the spin resistivity because it localizes the neutral fermions.

\end{document}